\documentclass[preprint,showpacs,preprintnumbers,amsmath,amssymb]{revtex4}

\usepackage{graphicx}
\usepackage{dcolumn}
\usepackage{bm}

\begin{document}

\preprint{ANL-HEP-PR-03-063}
\preprint{FERMILAB-Pub-03/234-T}
\preprint{DESY-03-036}

\title{Next-to-leading order QCD predictions for \\
$W+2$~jet and $Z+2$~jet production at the CERN LHC} 

\author{John Campbell}
\email{johnmc@hep.anl.gov}
\affiliation{
HEP Division, Argonne National Laboratory,\\
9700 South Cass Avenue, Argonne, IL 60439}

\author{R. K. Ellis}
\email{ellis@fnal.gov}
\affiliation{
Theory Group, Fermi National Accelerator Laboratory,\\
P.O. Box 500, Batavia, IL 60510}

\author{D. Rainwater}
\email{david.rainwater@desy.de}
\affiliation{
Theory Group, DESY \\
Notkestrasse 85, Lab. 2a, D-22603 Hamburg, Germany}

\today

\begin{abstract}
  We present cross sections and differential distributions for QCD
  radiative corrections to the QCD processes $pp\to W+\mbox{2~jets}$
  and $pp\to Z+\mbox{2~jets}$ at the CERN LHC.  Calculations are
  performed with the Monte Carlo program MCFM.  Cross section
  dependence on the renormalization and factorization scales is
  greatly reduced, except for the heavy-flavor case of $W^\pm
  b\bar{b}$, which has new features at next-to-leading order at the
  LHC.  We also present cross sections for $W^\pm b\bar{b}$ and
  $Z+\mbox{2~jets}$ in kinematic configurations relevant for Higgs
  boson searches.
\end{abstract}


\maketitle

\def\beq{\begin{equation}}
\def\eeq{\end{equation}}
\def\beqn{\begin{eqnarray}}
\def\eeqn{\end{eqnarray}}
\def\ran{\rangle}
\def\lan{\langle}
\def\bea{\begin{eqnarray}}
\def\eea{\end{eqnarray}}
\def\rar{\rightarrow}
\def\lar{\longrightarrow}
\def\lra{\leftrightarrow}
\def\ovl{\tilde}
\def\ie{{\it i.e.}}
\def\eg{{\it e.g.}}
\def\GeV{{\;\rm GeV}}
\def\MeV{{\;\rm MeV}}
\def\TeV{{\;\rm TeV}}
\def\blim{b_{\;\rm lim}}
\def\as{\alpha_s}
\def\bas{\bar{\alpha}_s}
\def\sla#1{\ifmmode%
\setbox0=\hbox{$#1$}%
\setbox1=\hbox to\wd0{\hss$/$\hss}\else%
\setbox0=\hbox{#1}%
\setbox1=\hbox to\wd0{\hss/\hss}\fi%
#1\hskip-\wd0\box1 }

\section{Introduction}

We report on the results of a calculation of the next-to-leading (NLO)
order QCD corrections to the QCD processes
\beqn
pp & \to & 
W^\pm\phantom{j}+\mbox{2~jets}
\to \ell^\pm\nu \phantom{j} + \mbox{2~jets} \; ,
\nonumber \\
pp & \to & 
Z/\gamma^* + \mbox{2~jets} \to \ell^+\ell^- + \mbox{2~jets} \; .
\eeqn
(single flavor, $\ell = e,\mu$ or $\tau$) at the CERN Large Hadron
Collider (LHC), \ie~$pp$ collisions at $\sqrt{s}=14$~TeV.  We ignore
electroweak (EW) contributions arising from \eg~$t$-channel
$W/Z/\gamma$ exchange between two scattered quarks, which are
typically ${\cal O}(1\%)$ for total rates.  Calculations for the
Fermilab Tevatron, $p\bar{p}$ collisions at $\sqrt{s}=2$~TeV, were
considered in a previous publication~\cite{Campbell:2002tg}, which
also introduced the calculational technique.  The results are made
with the Monte Carlo program MCFM, which makes full predictions for
any infra-red safe variable, including fully differential
distributions, for any set of experimental cuts, and any decay modes
of the $Z/\gamma^*$ intermediate states.  However, here we consider
only leptonic vector boson decays.  All leptons are treated as
massless, so the results treat $W,Z\to e,\mu,\tau$ flavors as
equivalent.

MCFM uses previously published matrix elements for the crossed
reactions $e^+e^-\to\mbox{4~partons}$~\cite{Bern:1997sc} and
$e^+e^-\to\mbox{5~partons}$~\cite{Nagy:1998bb}, in the four
dimensional helicity scheme.  The real corrections to the basic Born
processes, \ie
\beq
\mbox{parton} + \mbox{parton} \rightarrow W/Z/\gamma^* + \mbox{3~partons} 
\eeq
were published in
Refs.~\cite{Berends:1988yn,Hagiwara:1988pp,Nagy:1998bb}.  MCFM
incorporates these matrix elements using the subtraction
method~\cite{Ellis:1980wv,Catani:1996vz}.  These details and
those of the numerical checks performed may be found in
Ref.~\cite{Campbell:2002tg}.

As a useful comparison to more inclusive, well-studied weak boson
production, we simultaneously present MCFM results for the processes
$W,Z+\mbox{0,1~jets}$.  QCD corrections for these processes have been
known for some time~\cite{Giele:dj}, and have already proved
invaluable for Tevatron studies.  The inclusion of the $W/Z+1~{\rm
jet}$ processes in MCFM aided in understanding the issues to be faced
in implementing the more complicated $W/Z+2~{\rm jet}$ processes.  The
next-to-next-to-leading order (NNLO) corrections to total inclusive
$W,Z$ production are known~\cite{Hamberg:1990np}, but are not
implemented in MCFM.

We also present results where the additional jets are a heavy flavor
pair, specifically $b$ quarks, \ie~$W^\pm b\bar{b},Zb\bar{b}$, and
compare to the flavor-inclusive expectations.  This was previously
studied for the Tevatron~\cite{Ellis:1998fv,Campbell:2000bg}, but our
results are the first NLO predictions for these rates at the LHC.
These are of particular importance as backgrounds to a number of new
physics searches, especially for the Higgs boson.  Experimental
studies have so far had to rely on leading order (LO) predictions.
For the $W^\pm b\bar{b}$ processes we find a LO calculation
underestimates the rates at LHC energies by about a factor 2.4.

\section{General results}

Here we present results for the total observable cross sections for
the production of a vector boson and up to 2 jets.
Results specific to Higgs boson searches can be found in
Section~\ref{applications}.

\subsection{Input parameters}

For all our results we use the default set of EW parameters in MCFM,
which are given in Table~\ref{ewparams}.
\begin{table}
\begin{center}
\begin{tabular}{|c|c|c|c|} \hline
Parameter & Default value & Parameter & Default value \cr 
\hline
$M_Z$        & 91.187 GeV  & $\alpha(M_Z)$     & 1/128.89 \cr
$\Gamma_Z$   & 2.49 GeV    & $G_F$             & 1.16639$\times$10$^{-5}$ \cr
$M_W$        & 80.41 GeV   & $g^2_w$           & 0.42662 (calculated) \cr
$\Gamma_W$   & 2.06 GeV    & $\sin^2 \theta_w$ & 0.23012 (calculated) \cr 
\hline
\end{tabular}
\label{default} 
\end{center}
\caption{Default parameters in the program MCFM.
\label{ewparams}}
\end{table}
As noted in the table, some parameters 
are calculated using the effective field theory
approach~\cite{Georgi:1991ci},
\beq
e^2 =  4 \pi \alpha(M_Z),\;\; g_w^2 =  8 M^2_W  \frac{G_F}{\sqrt{2}},\;\;
\sin^2 \theta_w =  \frac{e^2}{g_w^2}.
\eeq
For simplicity we have taken the CKM matrix to be diagonal in the
$W+2~{\rm jets}$ process. As a consequence there are, for example, no
annihilating $u{\bar s}$ initial states for this case. This
approximation is not expected to influence any anticipated analyses.
For the other processes we retain only the Cabibbo sector of the CKM
matrix,
\beq
V_{CKM}= \left( \begin{array}{ccc}  
     0.975 & 0.222 & 0 \\ 
     0.222 & 0.975 & 0 \\ 
     0     & 0     & 1 \\ 
\end{array} \right) \; .
\eeq

The value of $\alpha_S(M_Z)$ is not adjustable; it is determined by the
chosen parton distribution.  A collection of modern parton distribution
functions (PDFs) is included with MCFM, but here we concentrate on the
CTEQ6L1 (LO) and CTEQ6M (NLO) sets~\cite{Pumplin:2002vw} which specify
$\alpha_S(M_Z)=0.130$ and $\alpha_S(M_Z)=0.118$ respectively.  Unless
otherwise stated, we take the factorization and renormalization scales
to be $\mu\equiv\mu_f=\mu_r=M_V$.

\subsection{Basic kinematic cuts and jet selection}
\label{cutsdesc}

For the general results presented here, we consider single leptonic
decays,
\beq
W^+ \to \nu e^+ \, , \qquad W^- \to \bar\nu e^- \, , 
\qquad Z/\gamma^* \to e^- e^+ \, . 
\eeq
$W^+$ and $W^-$ must be considered separately for $pp$ collisions, as
their rates are not equal, given the unequal quark/anti-quark parton
distributions.

Since these results are also intended to help calibrate experimentalists' Monte
Carlo results, the results must be obtained with kinematic cuts on the
final state particles for which they will be observable at the LHC
experiments ATLAS and CMS.  All leptons must satisfy
\beq
\label{eq:lep}
p_T(\ell) > 15~{\rm GeV}, \qquad |\eta_\ell| < 2.4,
\eeq
We also require that the charged dilepton invariant mass be greater
than $15~{\rm GeV}$.  This prevents the production of soft $e^- e^+$
pairs which would otherwise be copiously produced by the virtual
photon in the $Z/\gamma^*$ process.  We use the Run II $k_T$
clustering algorithm~\cite{Blazey:2000qt} to find jets, with a
pseudo-cone of size $R=0.4$. Jets are also subject to kinematic cuts
consistent with detector requirements for observability:
\beq
\label{eq:jet}
p_T(j) > 20~{\rm GeV}, \qquad |\eta_j| < 4.5.
\eeq
Finally, we require a minimum separation to isolate the leptons:
\beq
\label{eq:sep}
\triangle{R}_{\ell j} > 0.4 \; , \, \triangle{R}_{\ell\ell} > 0.2 \; .
\eeq
For the specific case of heavy-flavor jets, which are experimentally
observed via vertex tagging, the pseudo-rapidity restriction is more
severe:
\beq
|\eta_b| < 2.5 \; .
\eeq
We do not impose a cut on missing transverse momentum, $\sla{p}_T$,
for the $W^\pm +$~jets cases.  Not all analyses planned for the LHC
use such a cut, or use different values for it, depending on what else
is required in the event and machine luminosity.  Instead, we show the
differential distribution in this variable.

Except where noted, we consider only inclusive jets production:
$W,Z+n$~jets refers to the cross section for production of $W,Z$ and
$n$ or more jets.

\subsection{Scale dependence}
\label{scaledep}

The principal motivation for performing a NLO calculation is to reduce
the uncertainties in LO predictions.  In particular, any perturbative
prediction contains an unphysical dependence on renormalization and
factorization scales $\mu_r,\mu_f$.  The scale $\mu_f$ is introduced
during the factorization of the calculation into a perturbative hard
scattering part and non-perturbative PDFs.  The latter are taken as
input from data, with additional perturbative evolution. The
factorization and renormalization scales are often chosen to be equal,
$\mu\equiv\mu_r =\mu_f$, and we do this here.  Strictly speaking, the
two scales are unrelated and a calculation should be checked for
independence of each scale individually.  However, MCFM does not
currently allow for this possibility for most processes.
\begin{figure*}
\centering
\includegraphics[width=16cm]{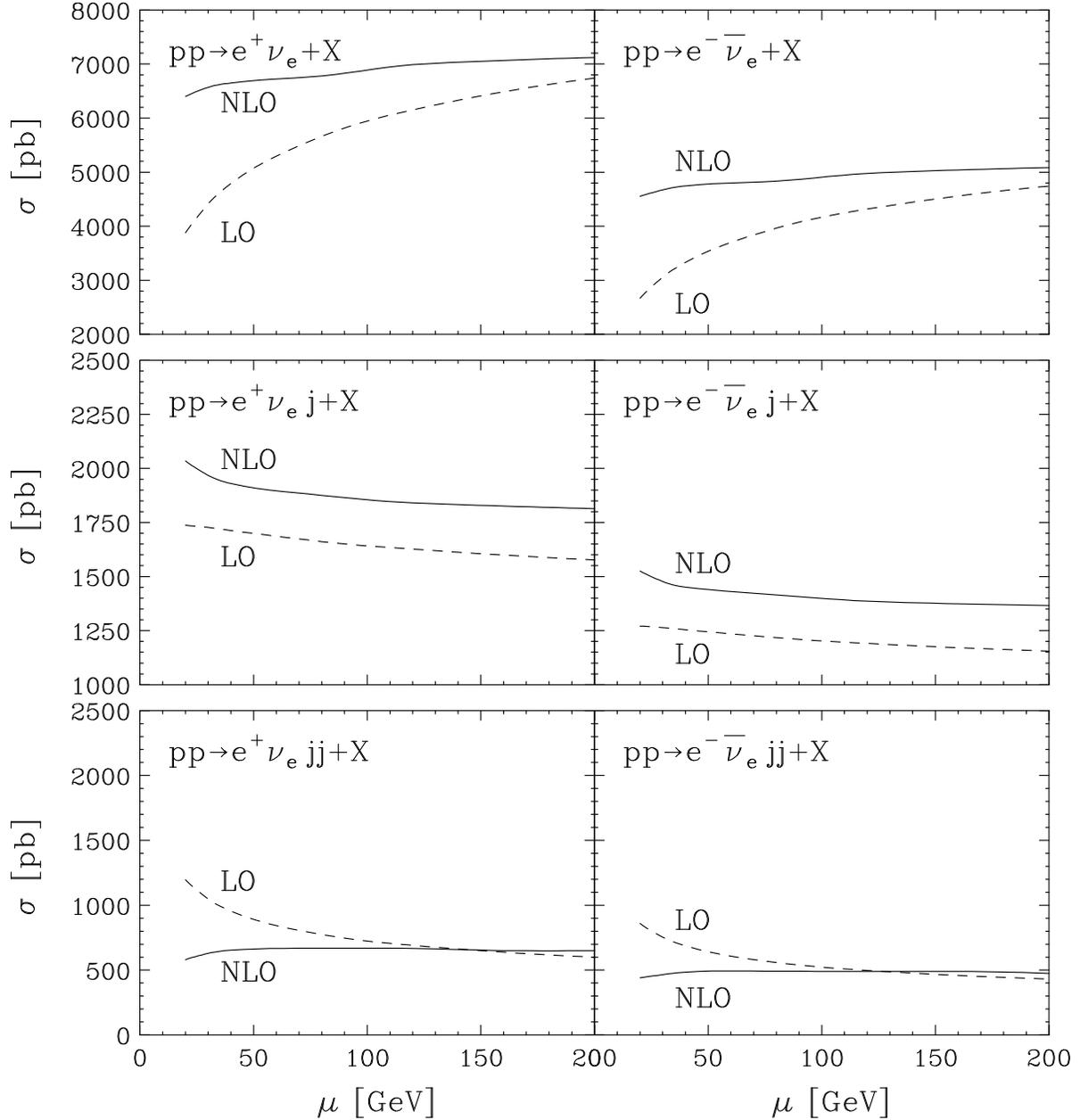}
\vspace{-5mm}
\caption{Scale dependence of the $e\,\nu_e +$~jets cross sections, 
  $\mu=\mu_r=\mu_f$, using the cuts described in
  Sec.~\protect{\ref{cutsdesc}}.  In general, $\mu$-dependence is
  significantly reduced at NLO compared to LO.}
\label{W.mudep}
\end{figure*}
\begin{figure*}
\centering
\centering
\includegraphics[width=16cm]{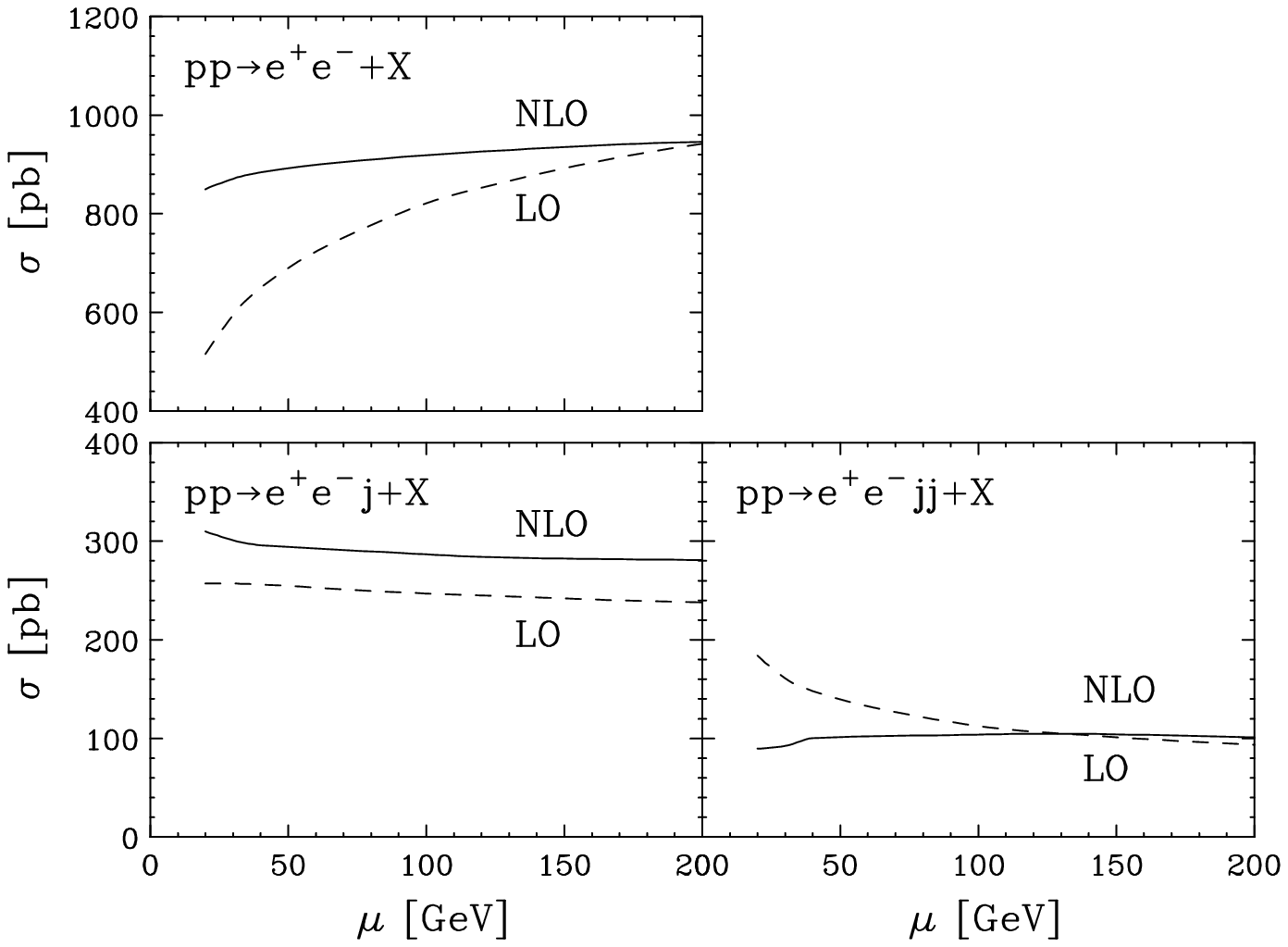}
\vspace{-5mm}
\caption{Scale dependence of the $e^+e^-j$ and $e^+e^-jj$ 
  cross sections, $\mu=\mu_r=\mu_f$, using the cuts described in
  Sec.~\protect{\ref{cutsdesc}}.}
\label{Z.mudep}
\end{figure*}

The magnitude of cross sections and the shape of differential
distributions can vary greatly between two different choices of scale.
This is often interpreted as an inherent ``theoretical uncertainty''.
Another strategy is to argue for a particular choice of scale, based
on the physics of the process under consideration.  Sometimes this
gives results at LO that are close to the NLO results, but is not
guaranteed.

A NLO calculation is an invaluable tool for investigating the issue of
scale dependence.  In a calculation performed through to order
$\alpha^n_s$, the residual scale dependence enters only at order
$\alpha^{n+1}_s$.  As a result, one expects that NLO predictions are
more stable under variations of the scale, $\mu$.  In addition, the
NLO result may provide further evidence to support a particular scale
choice that may have been deemed appropriate at leading order.  In
this case, the NLO result does not imply that the scale choice is
physically meaningful, but it can provide an easy method of
normalizing a LO calculation.  (A LO calculation typically proceeds
much more rapidly, and may need to be repeated multiple times for
different kinematic configurations.)
\begin{table}
\begin{center}
\begin{tabular}{|l|c|c|}
\hline
process & $\sigma_{LO}$ & $\sigma_{NLO}$ \\
\hline
$e^+\nu_e         +X$ & $5670$ & $6780^{+290}_{-130} $ \\
$e^-\bar\nu_e     +X$ & $3970$ & $4830^{+210}_{-90} $ \\
$e^+e^-           +X$ &  $803$ &  $915 \pm 31 $ \\
$e^+\nu_e\,j      +X$ & $1660$ & $1880^{+60}_{-50} $ \\
$e^-\bar\nu_e\,j  +X$ & $1220$ & $1420 \pm 40 $ \\
$e^+e^-\,j        +X$ &  $248$ &  $288^{+8}_{-7} $ \\
$e^+\nu_e\,jj     +X$ &  $773$ &  $669^{+0}_{-18} $ \\
$e^-\bar\nu_e\,jj +X$ &  $558$ &  $491^{+0}_{-7} $ \\
$e^+e^-\,jj       +X$ &  $116$ &  $105^{+1}_{-5} $ \\
\hline
\end{tabular}
\end{center}
\caption{Summary of LO and NLO cross sections [pb] for $W/Z+0,1,2$~jets,
  including single leptonic decay of the weak boson. The central value
  at NLO is for $\mu = M_V$, and the uncertainty at NLO is for scale
  variation from $\mu = M_V/2$ to $\mu = 2M_V$. No uncertainties are 
  shown for LO results, as they are all considerably larger than at NLO. 
  All Monte Carlo statistical uncertainties are less than $1\%$.}
\label{tab:NLO}
\end{table}

To show the improvement achieved in the present calculation, we plot
the scale dependence of the total inclusive $W,Z+0,1,2$~jets cross
sections, including leptonic decay, with the kinematic cuts of
Eqs.~(\ref{eq:lep}-\ref{eq:sep}), in
Figs.~\ref{W.mudep},\ref{Z.mudep}.  The NLO predictions for
$W,Z+0,1$~jets have been known for some
time~\cite{Arnold:1989dp,Arnold:1989ub,Giele:dj}, but we recalculate
them here with MCFM.  The $W+2$~jets behavior, Fig.~\ref{W.mudep}, is
practically independent of $\mu$, as in the $W,Z+0,1$~jets cases,
demonstrating a small theoretical uncertainty. Table~\ref{tab:NLO}
summarizes the NLO results for $V+$~jets processes, along with the
residual theoretical uncertainty due to scale dependence.  All Monte
Carlo statistical uncertainties are less than $1\%$.  While the QCD
corrections for $V+0,1$~jets are positive, for $V+2$~jets they are
slightly negative.

\subsection{Kinematic distributions}
\label{jetdists}

\begin{figure*}[h]
\centering
\includegraphics[width=16cm,height=12cm]{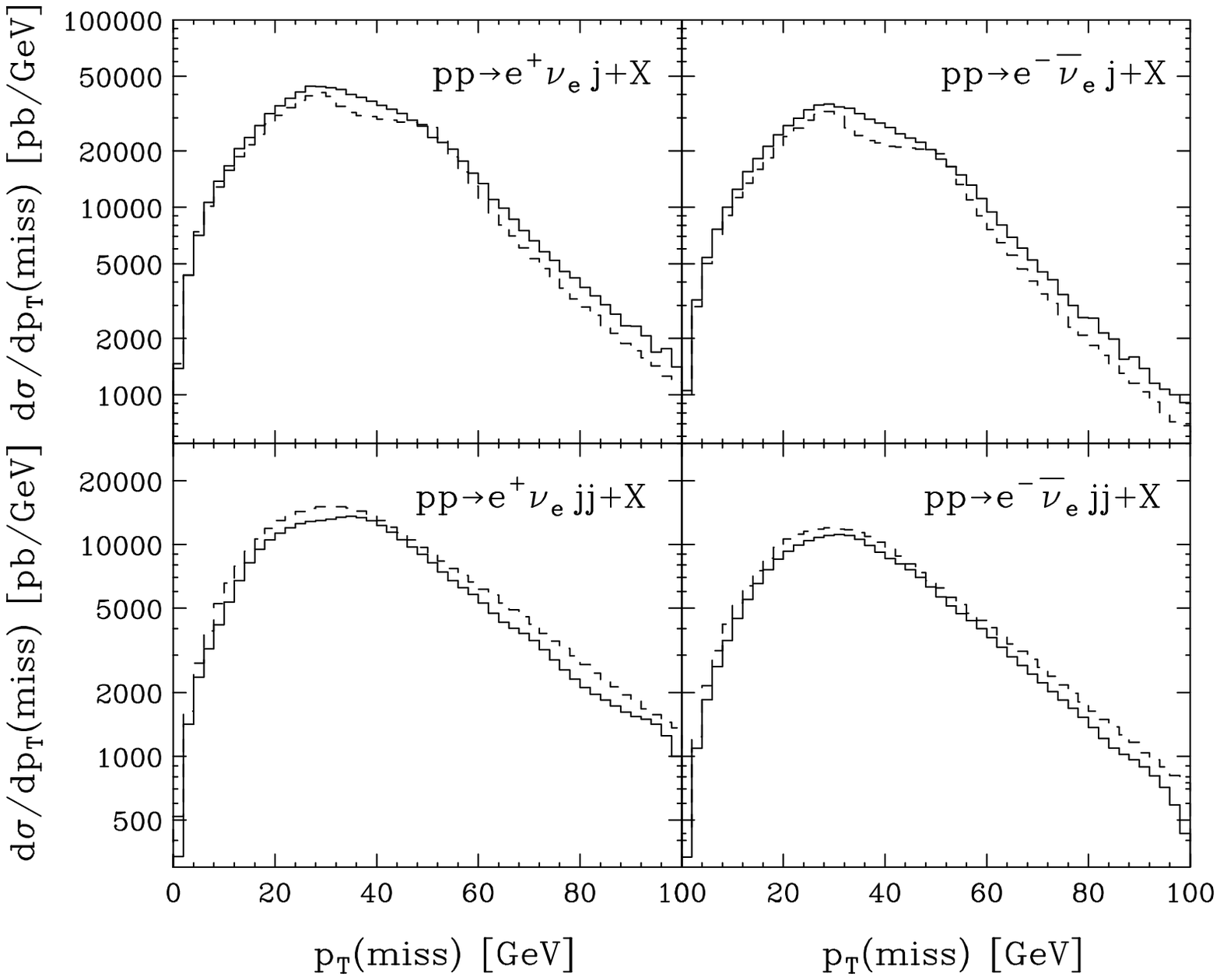}
\vspace{-5mm}
\caption{The missing transverse momentum distribution of $e^\pm\nu_e\,j$ 
  and $e^\pm\nu_e\,jj$ events at LO (dashed) and NLO (solid).}
\label{W.pTmiss}
\end{figure*}

\begin{figure*}[h]
\centering
\includegraphics[width=16cm,height=11cm]{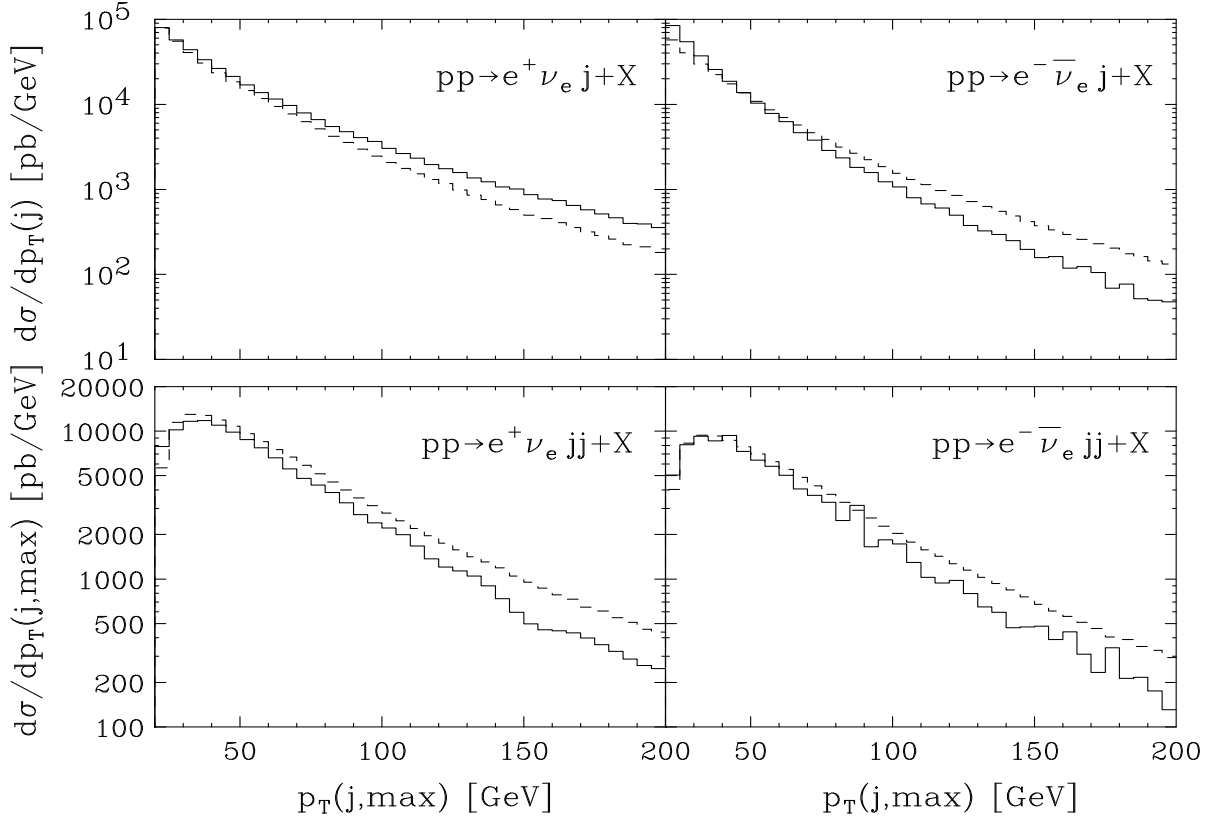}
\vspace{-5mm}
\caption{Hardest jet $p_T$ distribution of $e^\pm\nu_e\,j$ and 
  $e^\pm\nu_e\,jj$ events at LO (dashed) and NLO (solid).}
\label{W.pTjmax}
\end{figure*}

\begin{figure*}[ht]
\centering
\includegraphics[width=16cm]{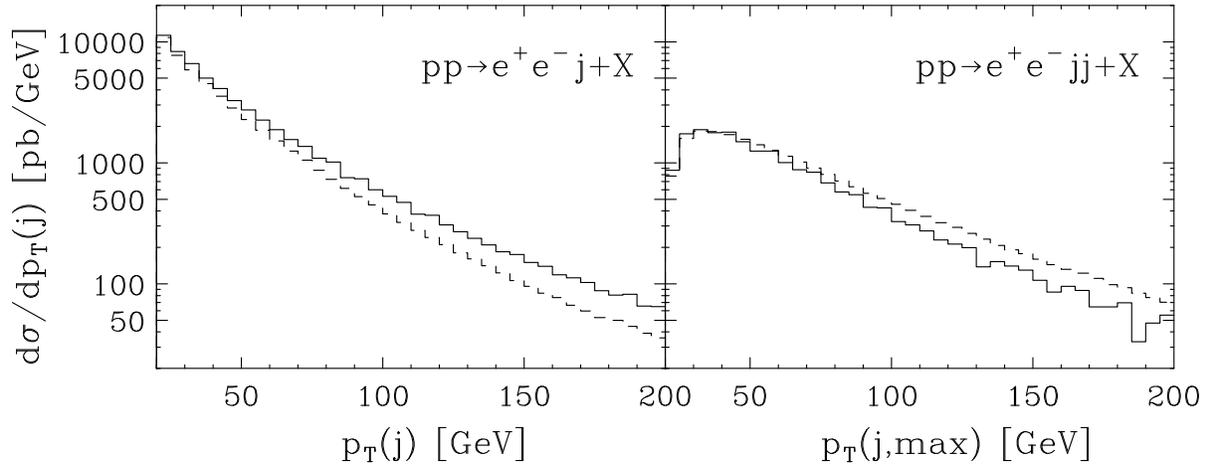}
\vspace{-5mm}
\caption{Hardest jet $p_T$ distribution in $e^+e^-\,j$ and $e^+e^-\,jj$ 
  events at LO (dashed) and NLO (solid).}
\label{Z.pTjmax}
\end{figure*}

Once again, we repeat some $W,Z+1$~jet results, in order to highlight
both the similarities and the differences with the corresponding
$2$-jet distributions.  We first show the missing transverse momentum
distributions for $W^\pm +$~jets in Fig.~\ref{W.pTmiss}, and then the
leading (hardest) jet $p_T$ distribution for the $W^\pm j$ and $W^\pm
jj$ cases in Fig.~\ref{W.pTjmax}, and for $Zj$ and $Zjj$ in
Fig.~\ref{Z.pTjmax}.
\begin{figure*}
\centering
\includegraphics[width=16cm]{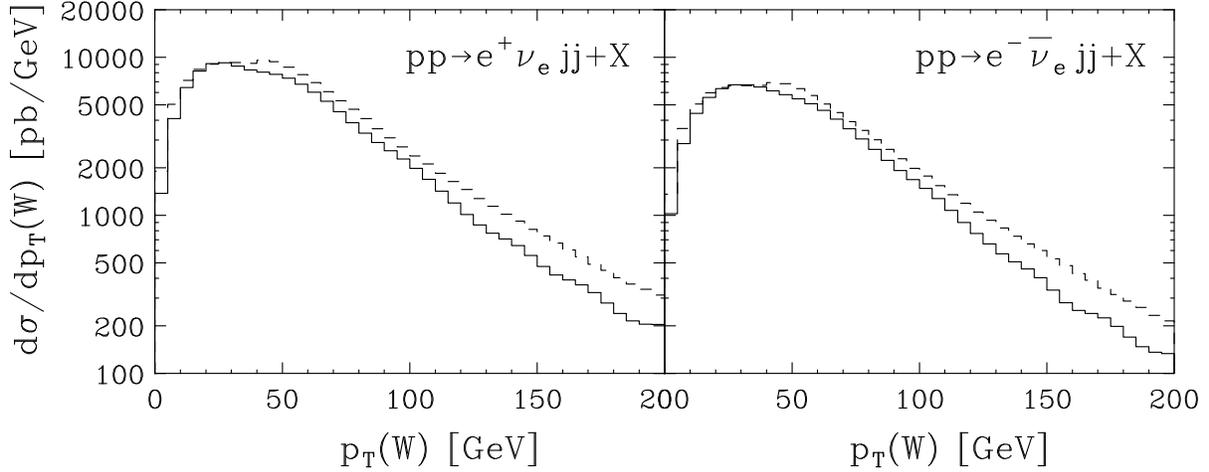}
\vspace{-5mm}
\caption{The boson $p_T$ distribution of $e^\pm\nu_e\,jj$  events at LO 
  (dashed) and NLO (solid).}
\label{W.pTV}
\end{figure*}
\begin{figure*}
\centering
\includegraphics[width=9cm]{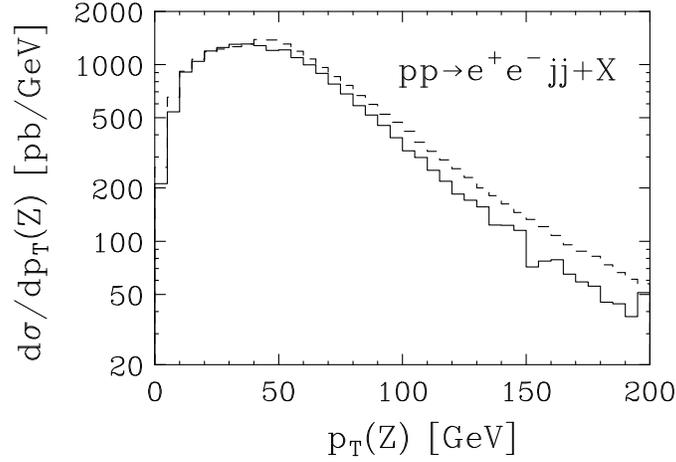}
\vspace{-5mm}
\caption{The boson $p_T$ distribution in $e^+e^-\,jj$ events at LO 
  (dashed) and NLO (solid).}
\label{Z.pTV}
\end{figure*}
\begin{figure*}
\centering
\includegraphics[width=16cm]{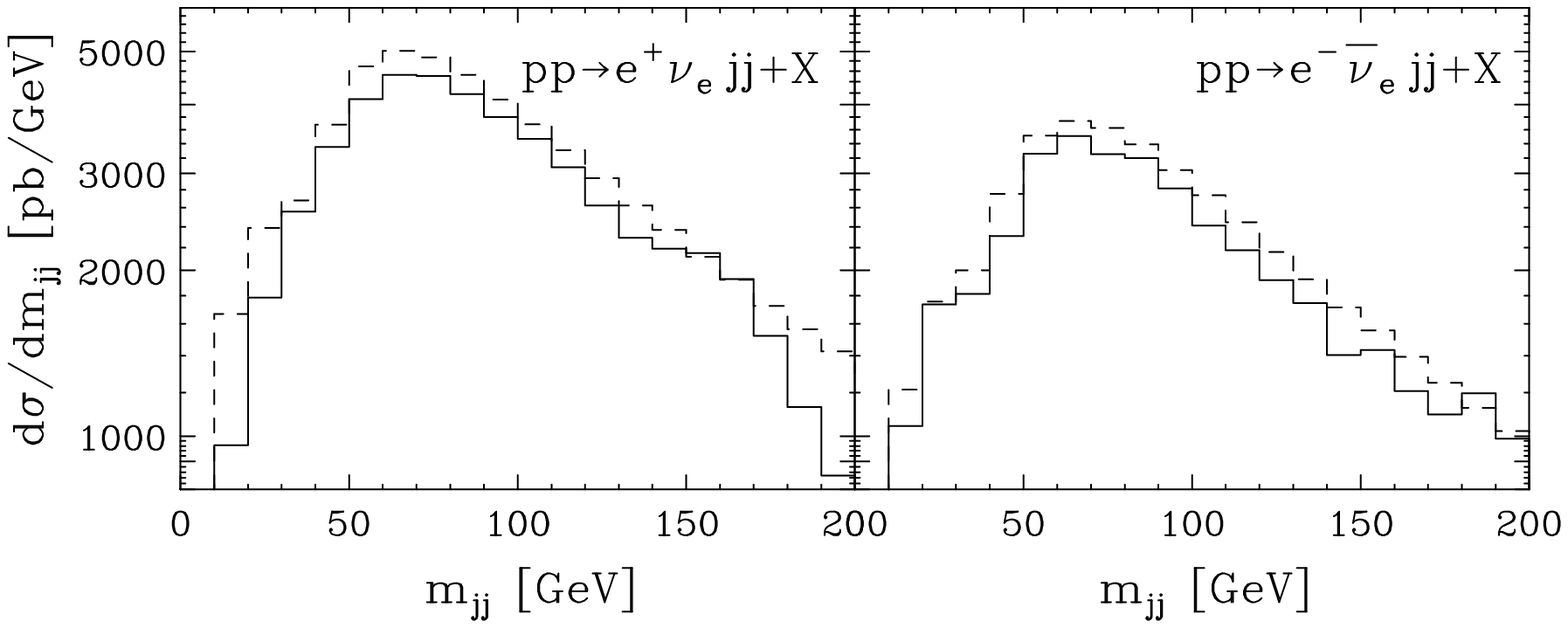}
\vspace{-5mm}
\caption{The jet pair invariant mass distribution of $e^\pm\nu_e\,jj$ 
  events at LO (dashed) and NLO (solid).}
\label{W.mjj}
\end{figure*}
\begin{figure*}
\centering
\includegraphics[width=8cm]{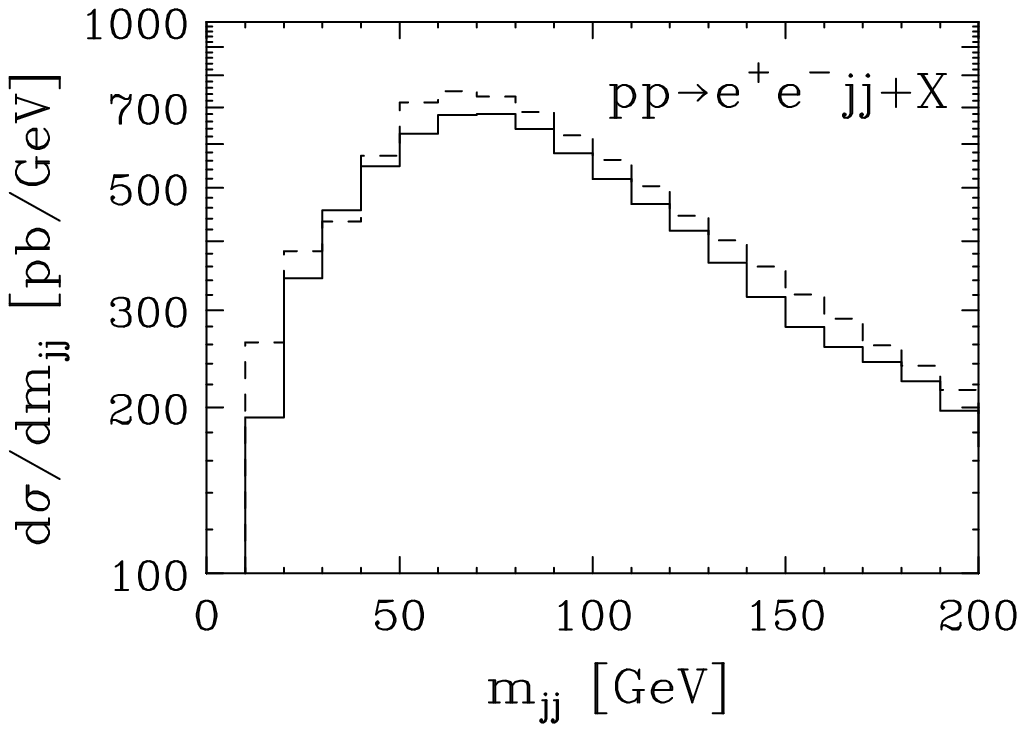}
\vspace{-5mm}
\caption{The jet pair invariant mass distribution of $e^+e^-\,jj$ 
  events at LO (dashed) and NLO (solid).}
\label{Z.mjj}
\end{figure*}
\begin{figure*}
\centering
\includegraphics[width=16cm]{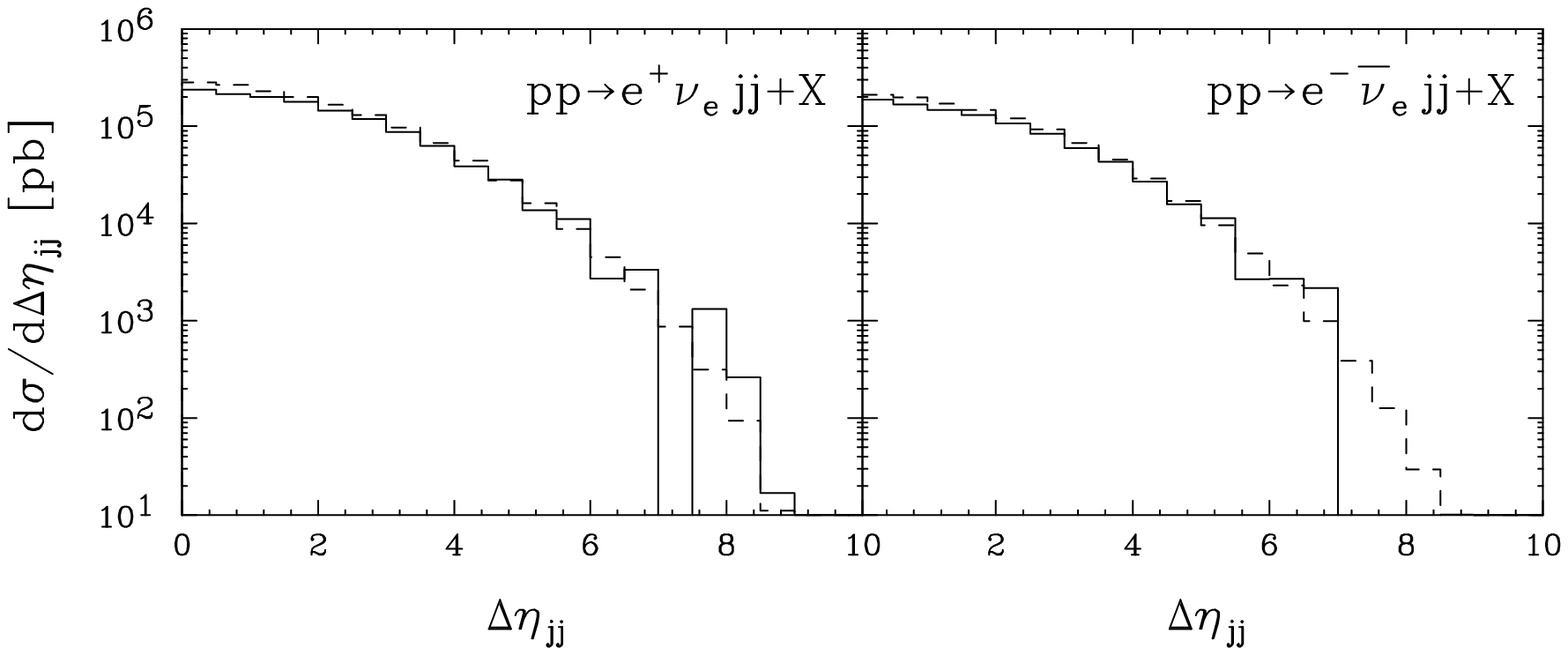}
\vspace{-5mm}
\caption{The jet pair $\eta$ separation distribution of $e^\pm\nu_e\,jj$
  events at LO (dashed) and NLO (solid).}
\label{W.dyjj}
\end{figure*}
\begin{figure*}
\centering
\includegraphics[width=8cm]{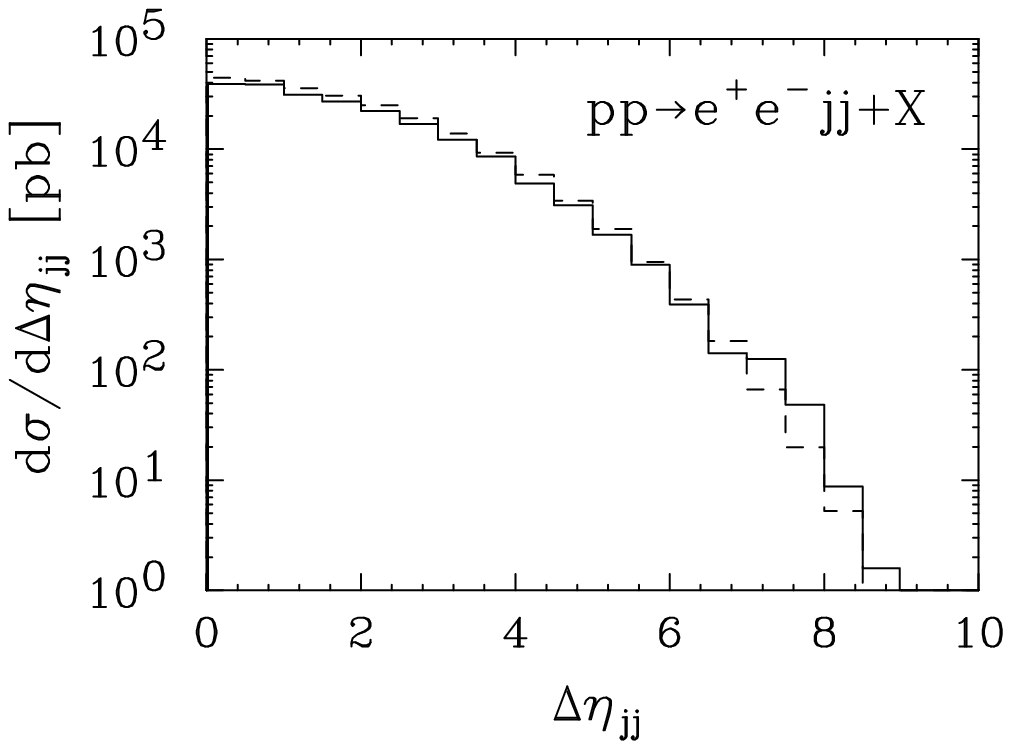}
\vspace{-5mm}
\caption{The jet pair $\eta$ separation distribution of $e^+e^-\,jj$ 
  events at LO (dashed) and NLO (solid).}
\label{Z.dyjj}
\end{figure*}
The leading order curves for $W^{\pm}+1$~jet in Fig.~\ref{W.pTmiss}
show a kinematic structure not present in the other curves.  At
leading order, the cut on the minimum $p_T$ of a jet,
Eq.~(\ref{eq:jet}) also imposes a minimum transverse momentum for the
$W$ boson, since at this order the transverse momenta are equal and
opposite.  Most events lie right above the cut, $p_T(W) \sim 20$~GeV.
The vector addition of this peaked distribution with the transverse
momentum generated in the decay leads to the structure shown in
Fig.~\ref{W.pTmiss}.  At NLO (or for higher jet multiplicity), the jet
transverse momentum cut does not impose the same constraint on the
momentum of the $W$ boson.  This explains the absence of the structure
in the other plots.  The existence of this structure in the LO plots
reinforces the importance of NLO calculations.

In contrast to the $W,Z+1$~jet cases, the $p_T$ distributions turn
over at small values for $V+2$~jets.  If the highest $p_T$ jet has a
transverse momentum, $p_T^{(1)}$, close to 20~GeV, there is little
phase space for the emission of a second, softer jet with
$p^{(1)}_T>p^{(2)}_T>p_T^{min}=20$~GeV.

We show the boson $p_T$ distributions in
Figs.~\ref{W.pTV},\ref{Z.pTV}, and the dijet invariant mass
distributions in Figs.~\ref{W.mjj},\ref{Z.mjj}.  The former show a
pronounced NLO deficit relative to LO at large values of $p_T(V)$,
whereas for the latter there is almost no change in shape from LO to
NLO.  The same holds true for the pseudo-rapidity separation of the
jet pair, as shown in Figs.~\ref{W.dyjj},~\ref{Z.dyjj}.

\section{Heavy flavor content of jets}
\label{heavy}

We calculate the fraction of $W,Z+2$~jet events that contain two heavy
quark jets, limiting ourselves to $b$ quarks, because they can be
tagged with high efficiency in experiment.  These processes are of
particular interest as backgrounds to new physics searches, in
particular Higgs bosons.  The calculations are identical to those
presented in Ref.~\cite{Campbell:2002tg}, but for $pp$ collisions at
the LHC energy of $\sqrt{s}=14$~TeV.  We work in the approximation in
which the $b$ quarks are taken to be massless, as the massive results
are not yet known at NLO, and we ignore contributions from processes
in which there are $b$ quarks already present in the initial state.

\begin{figure*}
\includegraphics[width=16cm]{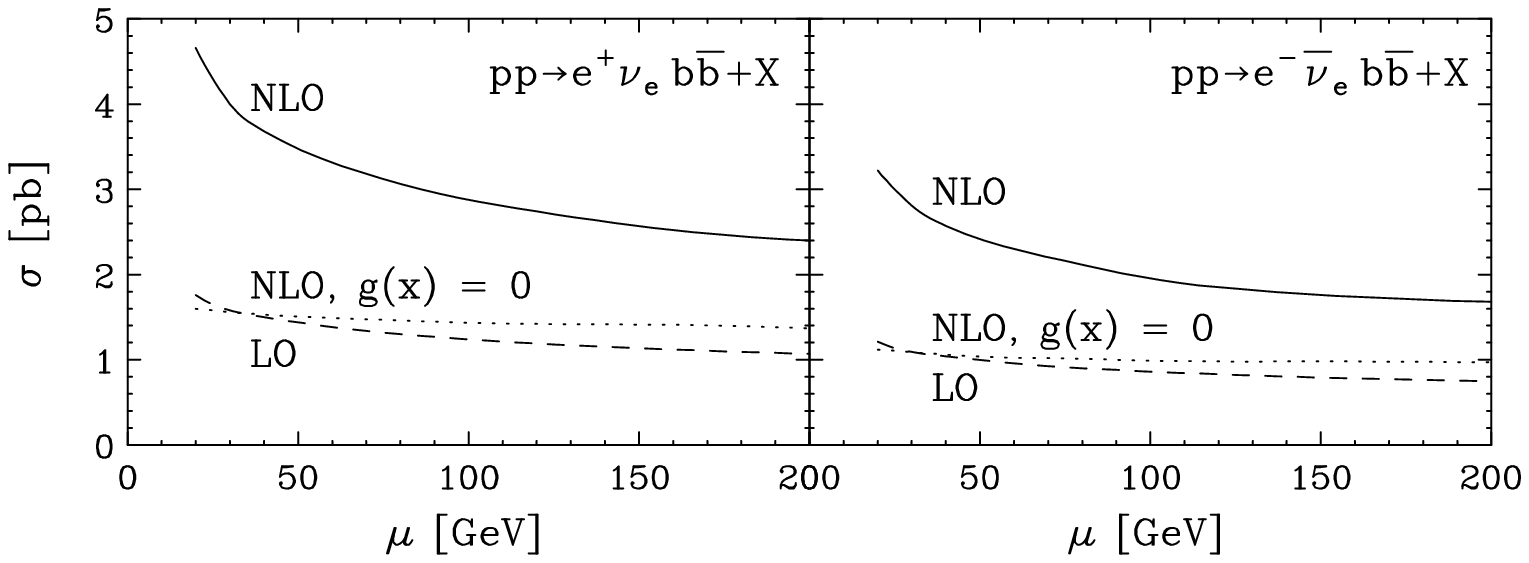}
\vspace{-5mm}
\caption{Scale dependence of the $e^\pm\nu_e\,b\bar{b}$ cross sections, 
  $\mu=\mu_r=\mu_f$, using the cuts described in
  Sec.~\protect{\ref{cutsdesc}}. Note the behavior of $Wb\bar{b}$ 
  at NLO: the NLO rate with initial gluon density set to zero is
  well-behaved, but with real initial gluons it behaves much more 
  like a LO calculation.  See text for additional comments.}
\label{Wbb.mudep}
\end{figure*}

\begin{figure*}
\includegraphics[width=9cm]{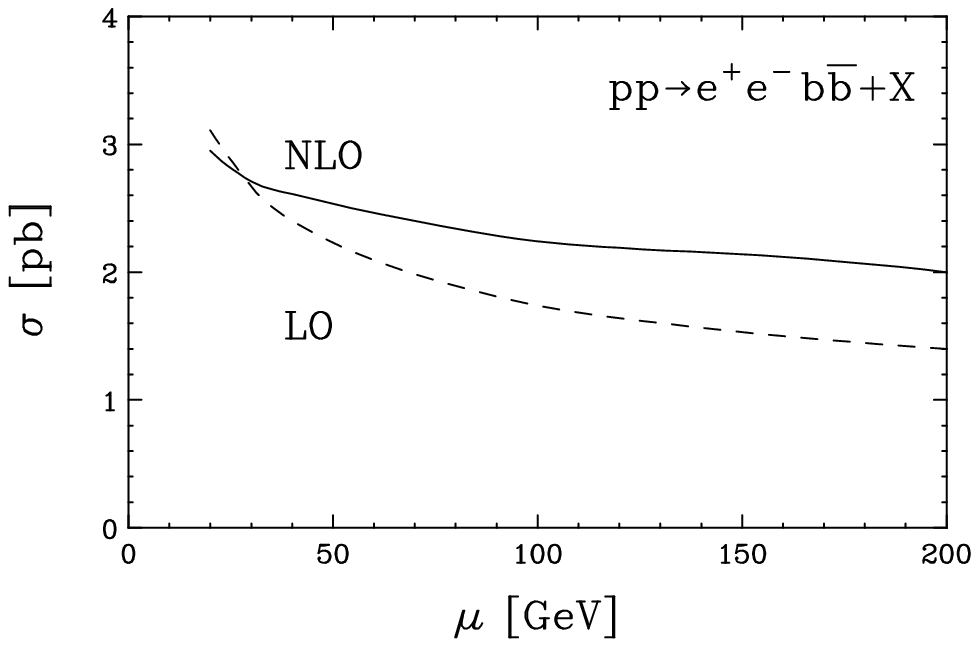}
\vspace{-5mm}
\caption{Scale dependence of the $e^+e^-\,b\bar{b}$ cross section, 
  $\mu=\mu_r=\mu_f$, using the cuts described in
  Sec.~\protect{\ref{cutsdesc}}.  The $Zb\bar{b}$ result is better
  behaved than in the case of $Wb\bar{b}$, but still retains an
  uncertainty of order $\pm 15\%$.}
\label{Zbb.mudep}
\end{figure*}

\begin{figure*}
\includegraphics[width=14cm]{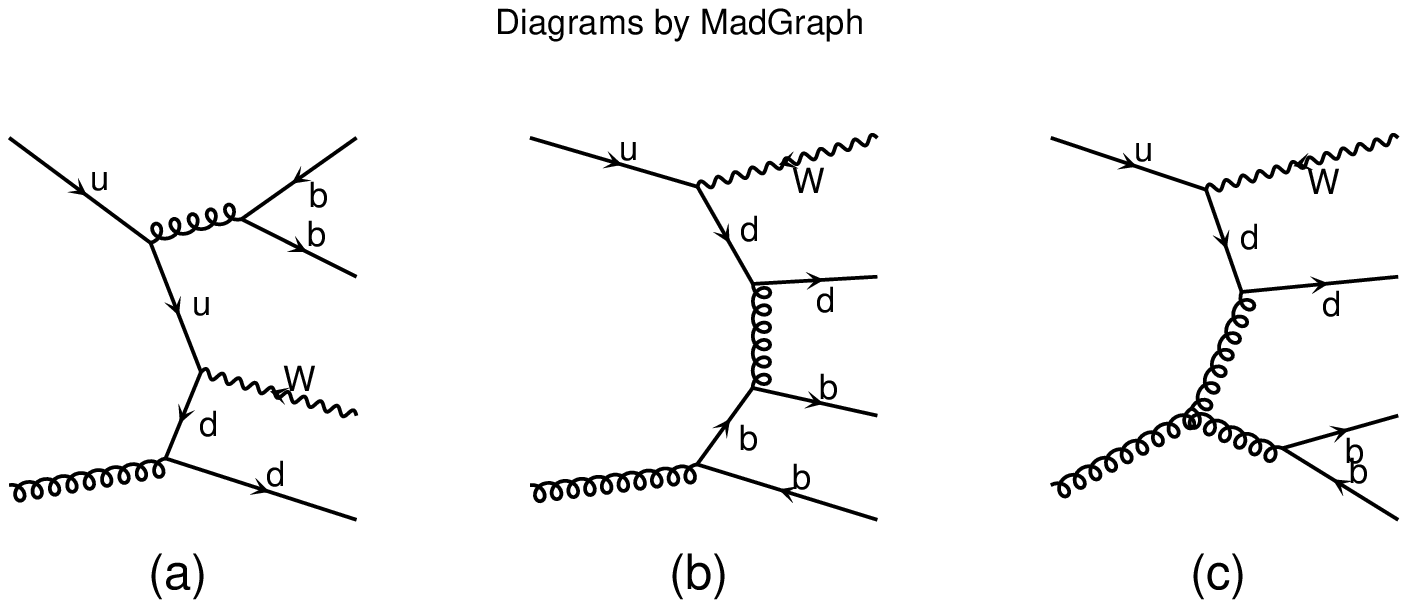}
\vspace{-5mm}
\caption{Some representative diagrams for the process 
  $ug\to dW^+b\bar{b}$. See text for discussion.}
\label{Wbbj}
\end{figure*}

We begin as in the non-heavy flavor case, checking the theoretical
uncertainty of the calculation by plotting the scale dependence of the
cross sections, shown in Figs.~\ref{Wbb.mudep},\ref{Zbb.mudep}.
The $W+b\bar{b}$ cross section shows much greater $\mu$ dependence at
NLO than at LO, contrary to na\"ive expectations.  For the central
scale choice $\mu=M_W$, the NLO result is a factor 2.4 larger than at
LO, and requires some explanation.  At LO, charge conservation allows
only $q\bar{q}$ initial states to contribute to $Wb\bar{b}$
production.  At NLO, the same is true for the virtual corrections, but
not for the real emission corrections, which contain subprocesses
$qg\to q'Wb\bar{b}$.  Some representative Feynman diagrams are shown
in Fig.~\ref{Wbbj}.  This allows the very large gluon luminosity at
the LHC to contribute.  However, if we ignore the additional
$b\bar{b}$ pair, then NLO corrections are of Drell-Yan type,
Fig.~\ref{Wbbj}(a), and might be expected to receive an additional
contribution of only $30\%$ at LHC energies (see Table~\ref{tab:NLO}).
\begin{figure*}[ht]
\centering
\includegraphics[width=16cm]{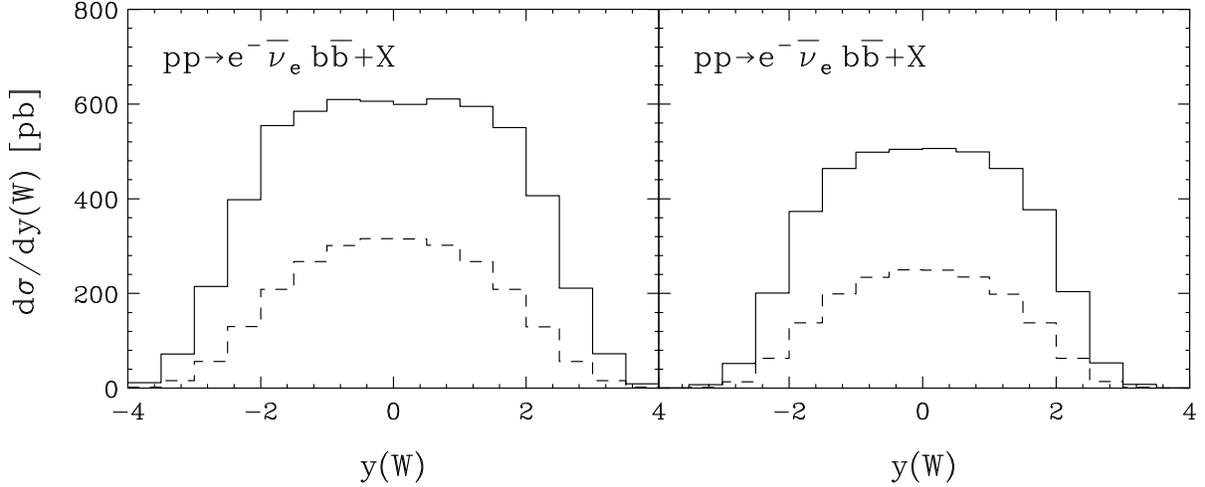}
\vspace{-5mm}
\caption{The $W$ rapidity distribution of $e^\pm\nu_e\,b\bar{b}$ events, 
  at LO (dashed) and NLO (solid). The broad plateau which appears at 
  NLO is indicative of the significant contribution from a $qg$-initiated
  component.}
\label{Wbb.yV}
\end{figure*}
The remaining contribution is due to diagrams of the type shown in
Figs.~\ref{Wbbj}(b,c), which contain no singularities for a finite $b$
mass; $W$ emission requires the massless quark propagator to be off
mass shell~\cite{footnote1}.  (We calculate with $m_b=0$, but avoid
the massless singularity by imposing a finite $p_T$ cut on both $b$
quarks.)  The contributions from diagrams of the type in
Fig.~\ref{Wbbj}(b) open up additional phase space, since they are not
suppressed by an s-channel gluon propagator which occurs in diagrams
of the type shown in Fig.~\ref{Wbbj}(a).  Furthermore, contributions
from diagrams of the type in Fig.~\ref{Wbbj}(c) contain a large
Casimir factor from the three-gluon vertex.  Diagrams (b) and (c)
constitute new, LO contributions that cannot appear in Drell-Yan.
Note, however, that one cannot consider individual diagrams
separately, as they must all be included together at the amplitude
level to maintain gauge invariance.  We demonstrate that the
additional diagrams are the cause of the hugely enhanced cross section
by calculating the NLO $W^\pm b\bar{b}$ cross section with initial
gluons turned off, \ie~$g(x)=0$.  This result has practically no $\mu$
dependence, as shown by the dotted curves in Fig.~\ref{Wbb.mudep}.
This is further supported by examining the rapidity distribution of
the emitted $W$ boson, shown in Fig.~\ref{Wbb.yV}.  The significant
NLO enhancement at large rapidity, resulting in a noticeable plateau
in the distribution, demonstrates the large size of the $qg$-initiated
component: quarks on average occur at larger values of Feynman $x$
than gluons, resulting in an overall boost of the $W$ boson in the
rest frame of the detector.  Thus, the dominant component of inclusive
$W^\pm b\bar{b}$ production at LHC energies, the summed
gluon-initiated real emission subprocesses, still contains large LO
theoretical uncertainty.  This can be resolved only by performing a
NLO calculation of $W^\pm b\bar{b}j$.

This situation does not exist for $Zb\bar{b}$ production, because lack
of a flavor-changing constraint on any quark line means that initial
state gluons contribute already at LO.  This is readily apparent from
Fig.~\ref{Zbb.mudep}, which shows NLO corrections much closer to the
size expected for Drell-Yan processes, and a residual uncertainty due
to scale dependence of about $20\%$.  Note the significant feature of
the heavy flavor case, that
$\sigma(W^+b\bar{b})\gtrsim\sigma(Zb\bar{b})\gtrsim\sigma(W^-b\bar{b})$,
as opposed to $\sigma(W^+jj)\gtrsim\sigma(W^-jj)\gg\sigma(Zjj)$
for the flavor-inclusive cross sections.

\begin{figure*}
\centering
\includegraphics[width=16cm]{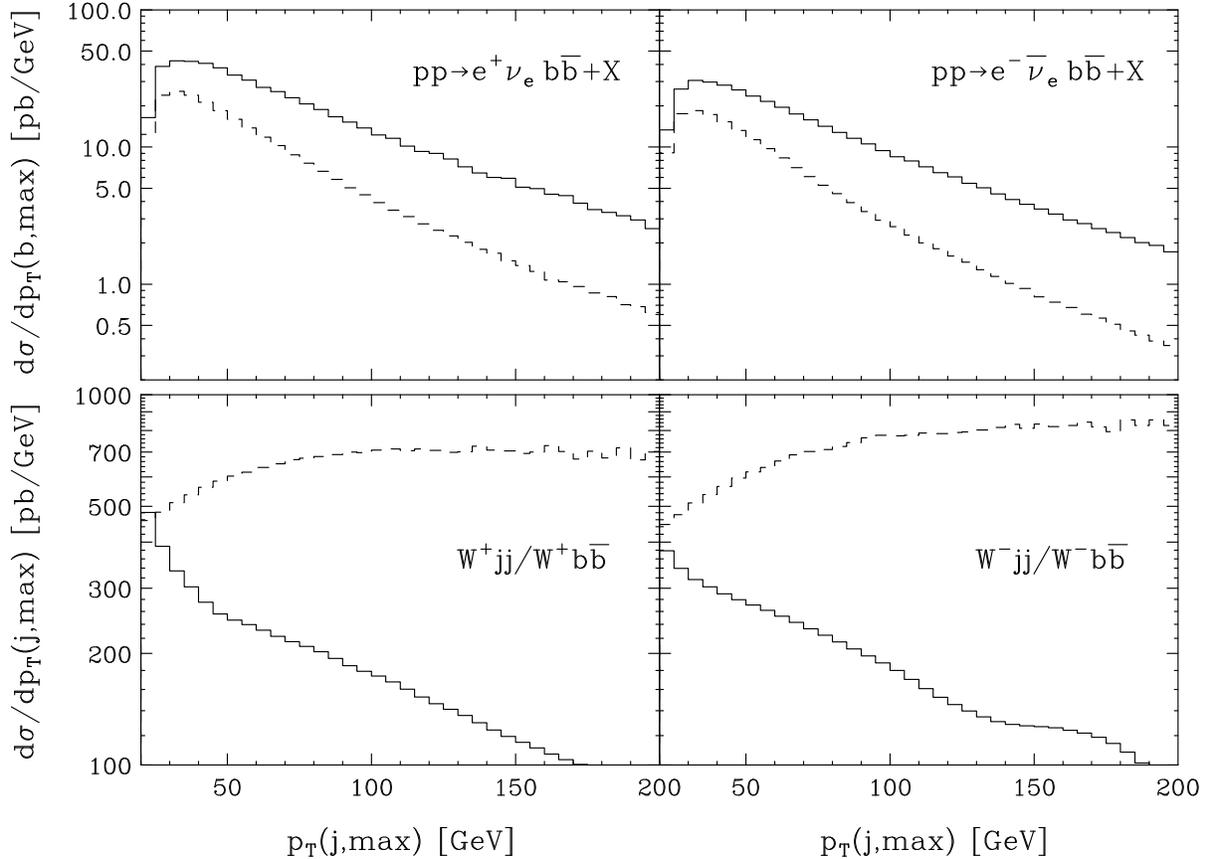}
\vspace{-5mm}
\caption{The leading jet $p_T$ distribution of $e^\pm\nu_e\,b\bar{b}$ 
  events, and of the ratio of $W^\pm\,jj$ to $W^\pm\,b\bar{b}$, at LO 
  (dashed) and NLO (solid). For the latter, the jets from $W^\pm\,jj$ 
  are also restricted to the region $|y_j|<2.5$, to match the 
  acceptance cuts that will be used for $W^\pm\,b\bar{b}$ events.}
\label{Wbb.pTjmax}
\end{figure*}
\begin{figure*}
\centering
\includegraphics[width=16cm]{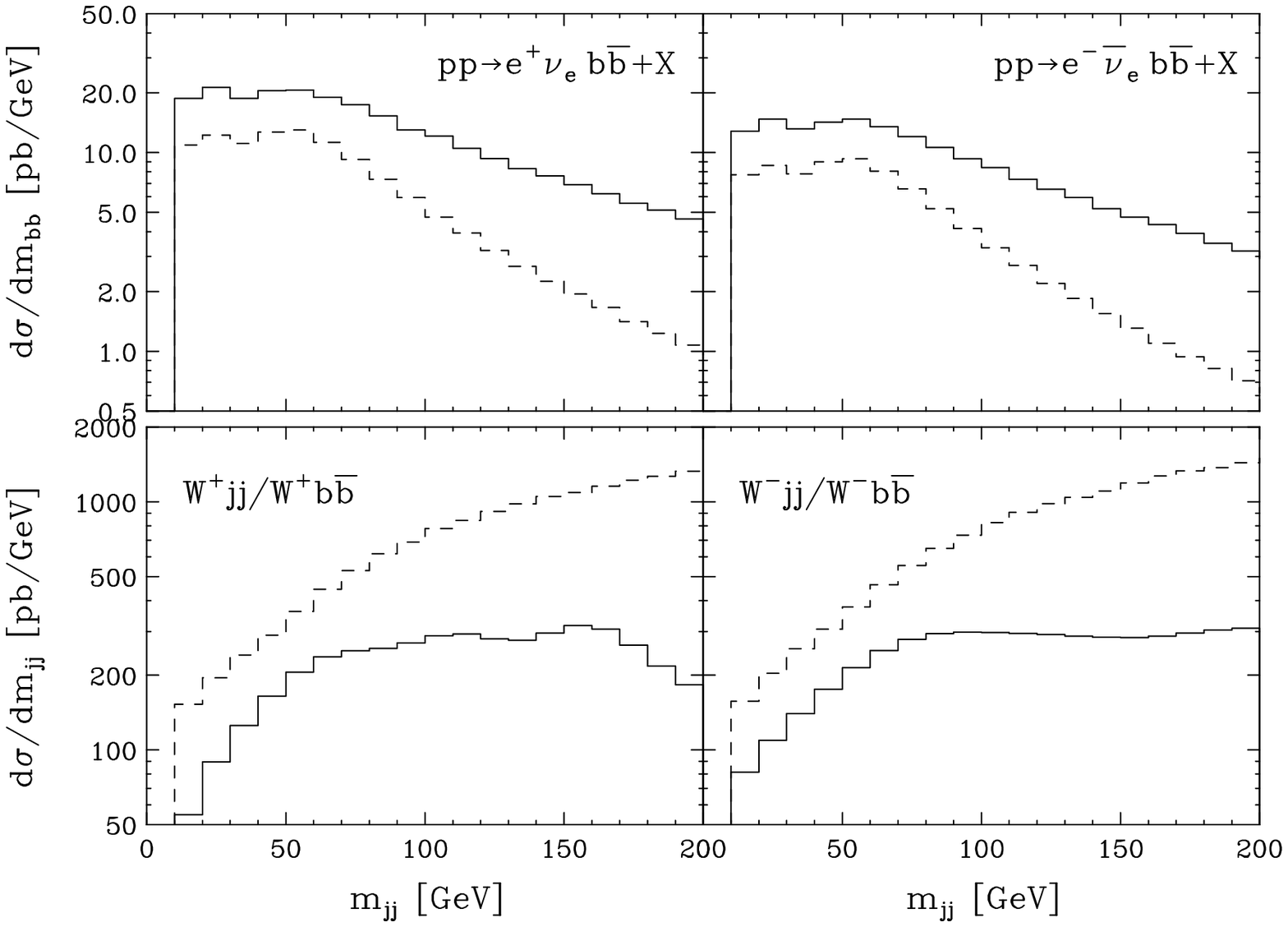}
\vspace{-5mm}
\caption{The dijet invariant mass distribution of $e^\pm\nu_e\,b\bar{b}$ 
  events, and of the ratio of $W^\pm\,jj$ to $W^\pm\,b\bar{b}$, at LO 
  (dashed) and NLO (solid). For the latter, the jets from $W^\pm\,jj$ 
  are also restricted to the region $|y_j|<2.5$, to match the 
  acceptance cuts that will be used for $W^\pm\,b\bar{b}$ events.}
\label{W.mbb}
\end{figure*}
As for the $W^\pm jj$ case in Sec.~\ref{jetdists}, we histogram the
differential $W^\pm b\bar{b}$ cross sections as a function of the $p_T$
of the leading jet, shown in Fig.~\ref{Wbb.pTjmax}, along with the
desired ratios of $W^\pm jj$ to $W^\pm b\bar{b}$ cross sections, to
illustrate the relative behavior of heavy flavor production.  In
principle, $W^\pm c\bar{c}$ production would behave similarly to $W^\pm
b\bar{b}$.  We also show the distributions as a function of dijet
invariant mass in Fig.~\ref{W.mbb}.  Including the $b$ quark mass at LO
is only a $4\%$ effect on the total cross section, but is contained
almost completely in the invariant mass range $2m_b<m_{bb}<20$~GeV. 
Consequently, the two $m_{bb}$ curves would exactly overlap each other
in Fig.~\ref{W.mbb}, except in the first bin.

\begin{figure*}
\centering
\includegraphics[width=16cm]{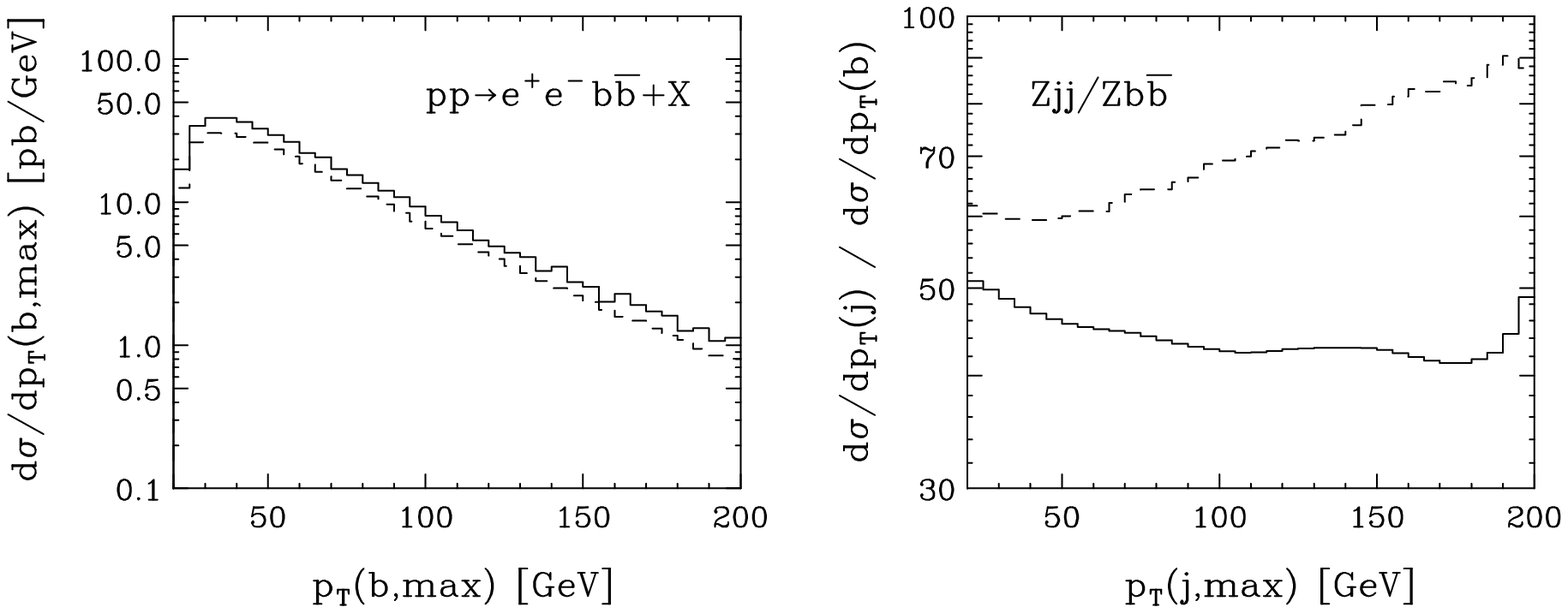}
\vspace{-5mm}
\caption{The leading jet $p_T$ distribution of $e^+e^-\,b\bar{b}$ 
  events,and of the ratio of $Zjj$ to $Zb\bar{b}$, at LO (dashed) and 
  NLO (solid). For the latter, the jets from $Zjj$ are also restricted 
  to the region $|y_j|<2.5$, to match the acceptance cuts that will 
  be used for $Zb\bar{b}$ events.}
\label{Zbb.pTjmax}
\end{figure*}

\begin{figure*}
\centering
\includegraphics[width=16cm]{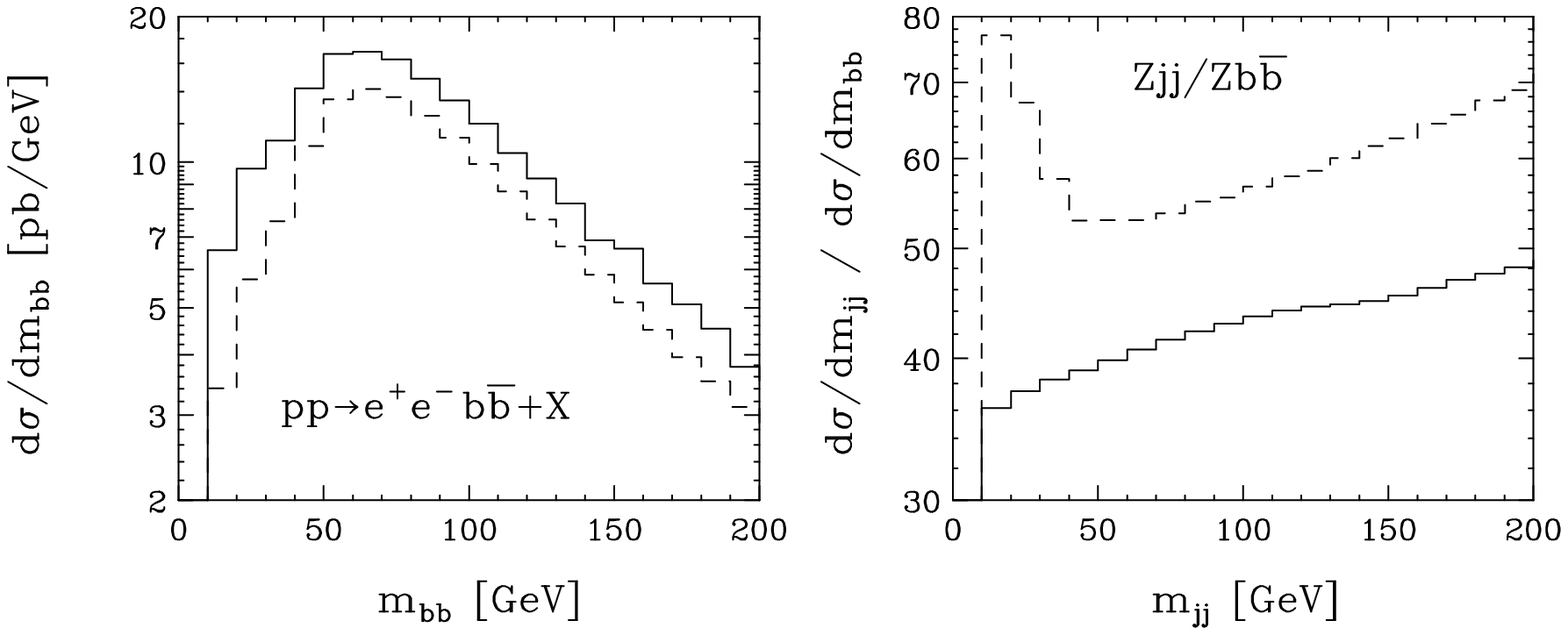}
\vspace{-5mm}
\caption{The dijet invariant mass distribution of $e^+e^-\,b\bar{b}$ 
  events, and of the ratio of $Zjj$ to $Zb\bar{b}$, at LO (dashed) 
  and NLO (solid). For the latter, the jets from $Zjj$ are also 
  restricted to the region $|y_j|<2.5$, to match the acceptance cuts 
  that will be used for $Zb\bar{b}$ events.}
\label{Z.mbb}
\end{figure*}

Similar histograms for the $Zb\bar{b}$ case are shown in
Figs.~\ref{Zbb.pTjmax},\ref{Z.mbb}. Table~\ref{tab:Vbb} summarizes the
$W^\pm,Z+b\bar{b}$ cross section results and scale uncertainties.
\begin{table}
\begin{center}
\begin{tabular}{|l|c|c|}
\hline
process & $\sigma_{LO}$ & $\sigma_{NLO}$ \\
\hline
$e^+\nu_e\,b\bar{b} +X$ &  $1.30^{+0.21}_{-0.18}$ &  $3.06^{+0.62}_{-0.54} $ \\
$e^-\nu_e\,b\bar{b} +X$ &  $0.90^{+0.14}_{-0.12}$ &  $2.11^{+0.46}_{-0.37} $ \\
$e^+e^-\,b\bar{b}   +X$ &  $1.80^{+0.60}_{-0.40}$ &  $2.28^{+0.32}_{-0.29} $ \\
\hline
\end{tabular}
\end{center}
\caption{Summary of LO and NLO cross sections [pb], including the 
  theoretical (scale) uncertainty at NLO, for $W/Z+b\bar{b}$, including
  leptonic decay of the weak boson. The central value at NLO is for 
  $\mu=M_V$. All Monte Carlo statistical uncertainties are less than $1\%$.}
\label{tab:Vbb}
\end{table}
%

\section{Applications}
\label{applications}

One of the principle goals of the LHC program is to determine the
nature of electroweak symmetry breaking (EWSB), which in the Standard
Model (SM) is due to a complex scalar $SU(2)$ doublet field.  Upon
acquiring a vacuum expectation value, it gives rise to both weak boson
and SM fermion masses and creates a single physical scalar, the Higgs
boson.  While there are other variations on this idea, or other models
which generate EWSB, nearly all of them have a physical scalar field
which couples to weak bosons and SM fermions similar to the SM Higgs
boson.  Many searches for it at hadron colliders look for associated
Higgs boson production, $pp\to W^\pm H, ZH$~\cite{VH.orig}, or a Higgs
boson produced between two far forward/backward high-$p_T$ jets.  The
latter is weak boson fusion (WBF) Higgs boson production, which arises
from a pair of initial-state quarks both emitting either a $W^+W^-$ or
$ZZ$ pair, which fuse to create the Higgs boson~\cite{Dicus:zg}.

Precise theoretical knowledge of the $W,Z+$~jets cross sections is of
particular interest at the LHC, since these processes constitute
significant backgrounds for the above Higgs boson searches, as well as
other SM processes of interest, such as top quark production, and many
other new physics searches.  We choose to examine only two particular
cases: $W^\pm b\bar{b}$ as a background to $W^\pm H;H\to b\bar{b}$,
and $Zjj$ as a background to WBF Higgs production, $pp\to Hjj$.

\subsection
{{\boldmath $W^\pm b\bar{b}$} and {\boldmath $W^\pm H;H\to b\bar{b}$}}

The most desirable Higgs boson search channels at the LHC are
inclusive production ($gg\to H$ via a top quark loop) or WBF Higgs
production, due to their much larger cross section and much cleaner
background environment, respectively.  Despite the smaller rate, the
$W^\pm H; H\to b\bar{b}$ channel has generated great interest because
of its potential to measure the bottom quark Yukawa coupling. Several
LHC experimental groups~\cite{Drollinger:2002uj,Kersevan:2002vu} have
studied the channel in detail.  These studies claim that the search
will be quite difficult for Higgs boson masses above the current LEP
limits, but possible.  While the signal is known at NLO, our
calculation of the $W^\pm b\bar{b}$ background at NLO is the first to
rectify this omission.

\begin{figure*}
\centering
\includegraphics[width=9cm]{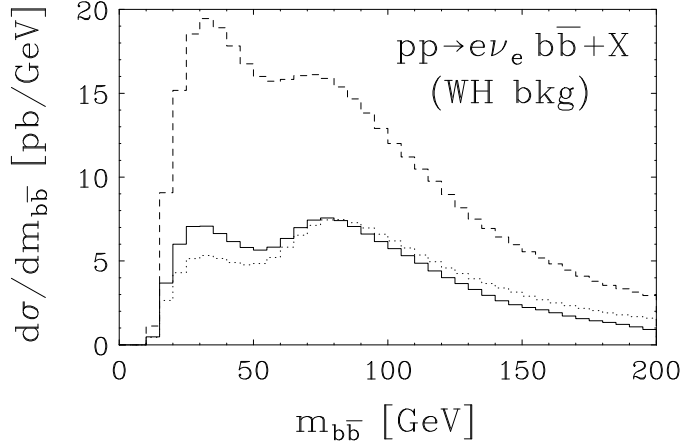}
\vspace{-5mm}
\caption{The $b\bar{b}$ invariant mass distribution of $e^\pm\nu_eb\bar{b}$ 
  events at LO (dotted), NLO (dashed), and NLO with a jet veto on the
  non-$b$ real emission (solid), with kinematic cuts specific to a
  potential $W^\pm H;H\to b\bar{b}$ search as described in the text.}
\label{Wbb.WHbkg}
\end{figure*}

As shown in Sec.~\ref{heavy}, the NLO cross section is more than two
times the LO result.  This does not bode well for this particular
Higgs boson search channel.  However, we must apply the same kinematic
cuts as in the current analyses and examine the NLO/LO ratio in a
dijet invariant mass bin relevant for the search.  We apply the
following cuts as in Ref.~\cite{Drollinger:2002uj}:
\beqn
p_T(\ell) > 20~{\rm GeV} \; , & |y_\ell| < 2.4 \; , \\
p_T(j) > 30~{\rm GeV} \; , & |y_j| < 2.5 \; , \\
\triangle{R}_{j\ell} > 0.4 \; , & \, \triangle{R}_{\ell\ell} > 0.2 \; ,
\eeqn
where the jet may or may not be a $b$ jet.  The resulting $b\bar{b}$
invariant mass distribution is shown for the summed $W^+$ and $W^-$
components in Fig.~\ref{Wbb.WHbkg}.  Assuming a Higgs boson mass of
$M_H=120$~GeV, and a mass bin of $\pm 20$~GeV,
\ie~$100<m_{jj}<140$~GeV, the NLO cross section is a factor 1.9 larger
than LO.

Since we know that the component causing the large enhancement
typically has an additional jet emission at large $p_T$, we impose a
non-$b$ jet veto at NLO in the real emission calculation. This turns
the cross section into an exclusive result, where the heavy flavor is
assumed to have been tagged. This NLO exclusive result is about $10\%$
smaller than the LO result, making the background much more
manageable.  {\it However}, we emphasize that use of this result
requires normalizing a parton shower calculation to this cross section
{\it after} a light-jet veto is imposed in the parton shower Monte
Carlo! The residual uncertainty from scale variation is $\approx
^{+44}_{-70}\%$.

\subsection{$Zjj$ and $Hjj$}
\label{WBF}

A characteristic feature of WBF Higgs boson production is $Hjj$ events
with the Higgs boson appearing at central rapidities, but the two
scattered quarks appearing at large rapidities with significant $p_T$.
Because the process is color singlet exchange, the events typically
have little central jet emission.  In contrast, QCD $Zjj$ production
prefers the jets to be much more central, while the $Z$ comes from
Bremsstrahlung and is at larger rapidites.  Also, the color triplet or
octet exchange gives rise to significant additional central jet
emission, known as ``minijet''
activity~\cite{Barger:1994zq,Rainwater:1996ud}.  We use MCFM here to
provide a NLO normalization of the QCD $Zjj$ rate in a region of phase
space relevant for WBF Higgs boson searches where the Higgs boson
decays to tau pairs.

Standard kinematic cuts for these searches will require the jets,
frequently called ``tagging jets'', to have a minimum separation of
about 4 units of pseudo-rapidity, to appear in opposite hemispheres,
and for the Higgs decay products (\eg~$\tau^+\tau^-$) to appear
between the jets~\cite{Rainwater:1998kj}.  In addition to the basic
cuts of Eqs.~(\ref{eq:lep}-\ref{eq:sep}), we thus impose the following
cuts:
\beq
|\eta^{(1)}_j - \eta^{(2)}_j| > 4.0 \;\; , \qquad
\eta^{(1)}_j\cdot\eta^{(2)}_j < 0   \;\; , \qquad
\eta^{min}_j + 0.4 < \eta_{\ell_1,\ell_2} < \eta^{max}_j - 0.4 \;\; ,
\eeq
where $\eta^{max}_j$ $(\eta^{min}_j)$ is the larger (smaller) of
$\{\eta^{(1)}_j,\eta^{(2)}_j\}$.  This reduces the QCD $Zjj$ cross
section from 94.9~pb to 2.9~pb at LO, and 104~pb to 3.4~pb at NLO.
\begin{figure*}
\centering
\includegraphics[width=9cm]{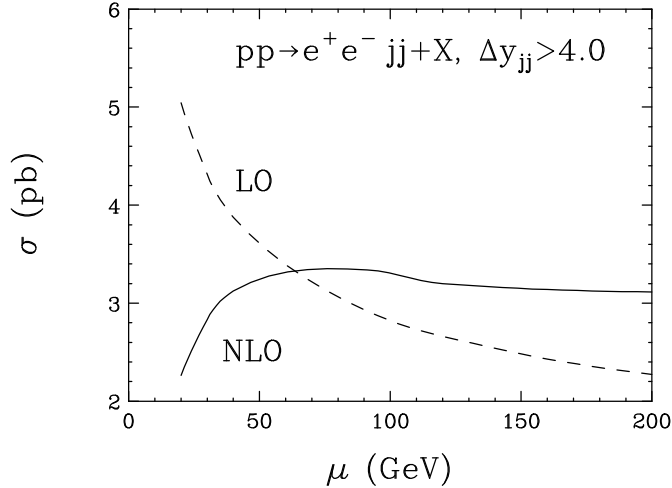}
\vspace{-5mm}
\caption{Scale dependence of the $e^+e^-jj$ ($\tau^+\tau^-jj$) cross
  section, $\mu=\mu_r=\mu_f$, using the cuts described in
  Sec.~\protect{\ref{WBF}} to examine the region of far forward
  tagging jets.}
\label{Z2j.gap.mudep}
\end{figure*}
Comparison to previous experimental studies are difficult, as all of
these used dynamical scales, and were all different from each
other~\cite{wbf_exp}.

\begin{figure*}
\centering
\includegraphics[width=16cm]{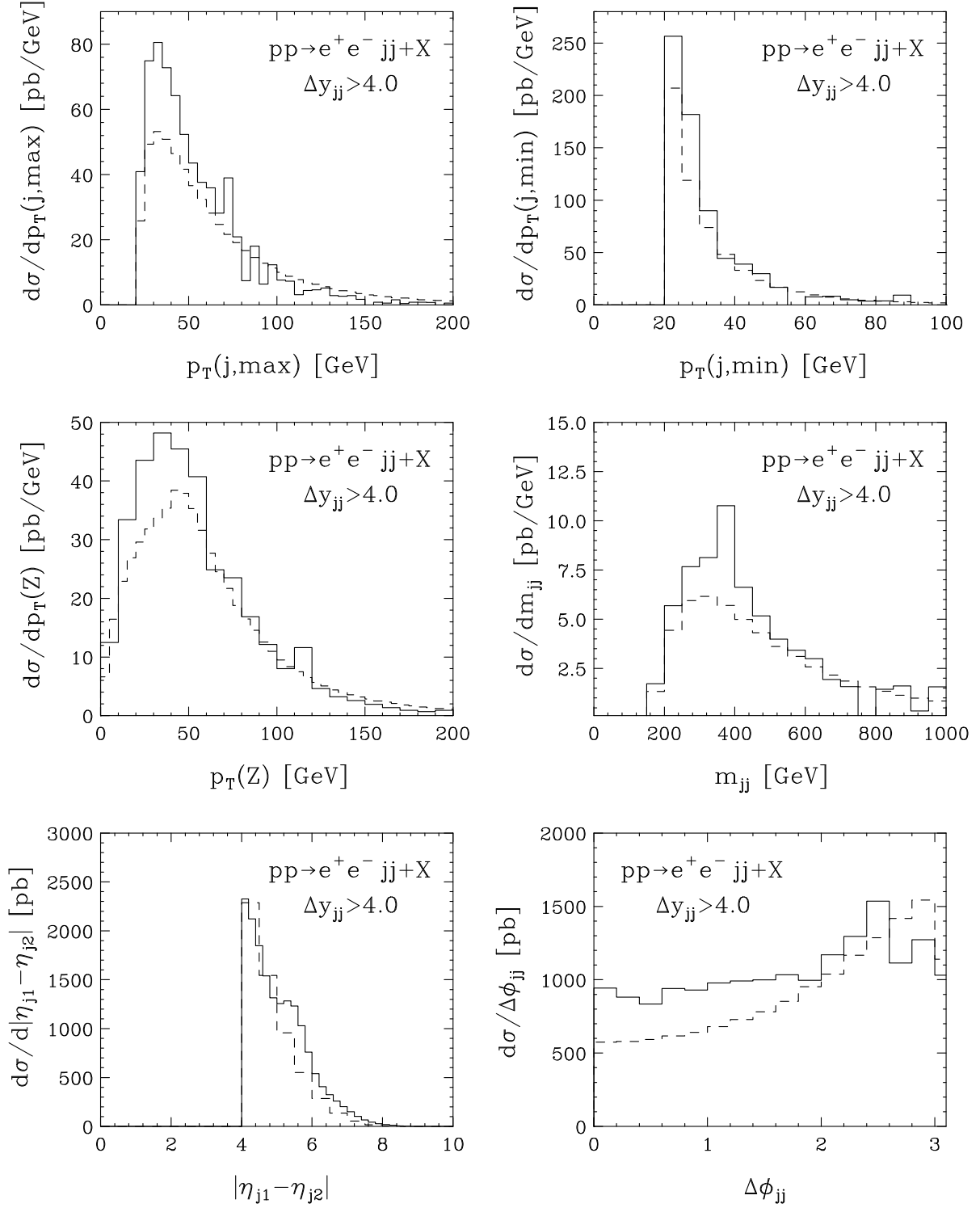}
\vspace{-5mm}
\caption{Various kinematic distributions for QCD $e^+e^-jj$
  ($\tau^+\tau^-jj$) production as a background to $Hjj$ production,
  as described in the text.}
\label{Z2j.gap}
\end{figure*}

We show the scale dependence for $Z+2$~jets in this phase space
configuration in Fig.~\ref{Z2j.gap.mudep}.  The NLO is quite stable,
as in the total cross section result, except for very low values of
$\mu$.  This is not uncommon in NLO calculations and typically
indicates the presence of logarithms that can become large with small
choices of $\mu$, but are not necessarily a problem at that order in
the perturbative expansion.  We also plot several differential
distributions in Fig.~\ref{Z2j.gap} to demonstrate that there do not
appear to be any large NLO corrections to the shapes in this small
region of phase space; the NLO distributions in general appear to be
very similar to the LO distributions.  One notable exception is the
azimuthal angle between the tagging jets, which is flatter at NLO,
compared to the preferential back-to-back configuration of the jets at
LO.  This is likely due to emission of final state radiation spreading
out the preference for the QCD jet pair to be back-to-back, but
requires further investigation.  This distribution is of special
relevance for studies of Higgs $CP$ transformation
properties~\cite{Plehn:2001nj}.

%

\section{Conclusions}

We have presented the first results for the implementation of $W/Z+2~$
jet production at NLO at the LHC, including the specific cases of
heavy flavor jets.  An analysis based on inclusive jet production for
the LHC shows that in most cases the usual benefits of NLO are
realized, among them a reduced scale dependence and hence an improved
normalization for kinematic distributions.  The outstanding exception
is the case of $W^\pm b\bar{b}$ production, which at one higher order
in the QCD coupling experiences additional LO contributions that
dramatically enhance the total rate.  We showed that this additional
contribution typically leads to an additional non-$b$ jet at large
$p_T$ and pseudo-rapidity, which can be identified in the detectors.

These processes are of particular interest at the LHC as backgrounds
to other SM processes and searches for new physics.  We analyzed two
cases in particular.  The first analysis, $W^\pm b\bar{b}$ as a
background to $WH;H\to b\bar{b}$ for a light Higgs boson, suffers
considerably from the unexpected large NLO corrections, but may be
controlled somewhat by vetoing non-$b$ jets.  This will require
additional analysis with parton shower Monte Carlo normalized to our
NLO jet-vetoed rates, after the veto of additional jets radiated in
the parton shower.  The second analysis, $Zjj$ as a background to weak
boson fusion Higgs production, $Hjj$, shows that the overall
corrections are not very large for a central scale choice of
$\mu=M_Z$, only about $+14\%$.  However, most $Hjj$ analyses have used
dynamical scale choices of $p_T(j,min)$, which average only a few tens
of GeV, so these analyses will generally benefit from a reduction in
the QCD $Zjj$ background normalization.  In addition, we have shown
that no large NLO corrections appear in this tiny region of phase
space with jets widely separated in pseudo-rapidity, and various
kinematic differential distributions at NLO appear very much like
their LO counterparts.  This allows simple overall normalization of
parton shower Monte Carlo based on the total rates, but does not
address the issue of vetoing events with additional minijet activity.

\begin{acknowledgments}
  This work was supported in part by the U.S. Department of Energy
  under Contracts No. W-31-109-ENG-38 (Argonne) and No.
  DE-AC02-76CH03000 (Fermilab).  We would like to thank Volker
  Drollinger, Tilman Plehn and Dieter Zeppenfeld for useful 
  discussions.
\end{acknowledgments}



\bibliographystyle{plain}


\clearpage

\end{document}